\documentclass[twocolumn,showpacs,preprintnumbers,amsmath,amssymb,pra]{revtex4}

\usepackage{epsfig}
\usepackage{graphics}
\usepackage{graphicx}
\usepackage{bm}
\usepackage[dvips]{color}

\newcommand{\an}[1]{\hat{#1}}
\newcommand{\ad}[1]{\hat{#1}^{\dag}}
\newcommand{\be}{\begin{equation}}
\newcommand{\ee}{\end{equation}}
\newcommand{\bea}{\begin{eqnarray}}
\newcommand{\eea}{\end{eqnarray}}
\newcommand{\ket}[1]{| #1 \rangle}
\newcommand{\bra}[1]{\langle #1 |}

\newcommand{\medial}[1]{\langle #1 \rangle}

\begin{document}

\title{Quasi-deterministic generation of maximally entangled states of \\
two mesoscopic atomic ensembles by adiabatic quantum feedback}
\author{Antonio Di Lisi$^{1}$, Silvio De Siena$^{1}$, Fabrizio Illuminati$^{1}$, and David Vitali$^{2}$}
\affiliation{$^{1}$ Dipartimento di Fisica ``E. R. Caianiello'', Universit\`{a} di Salerno, INFM - Coherentia, and \\
INFN Sezione di Napoli, Gruppo collegato di Salerno, Via S. Allende, I-84081 Baronissi (SA), Italy \\
$^{2}$ Dipartimento di Fisica, Universit\`a di Camerino, I-62032 Camerino (MC), Italy}


\begin{abstract}
We introduce an efficient,
quasi-deterministic scheme to generate maximally entangled states
of two atomic ensembles. The scheme is based on quantum non-demolition
measurements
of total atomic populations and on adiabatic
quantum feedback conditioned by the measurements
outputs. The high efficiency of the scheme is tested and confirmed
numerically for ideal
photo-detection as well as in the presence of losses.
\end{abstract}


\maketitle

\section{Introduction}

Quantum entanglement is one of the most fundamental aspects of quantum
mechanics, as well as an essential resource in quantum communication
and information processing.
Although very difficult to realize, entangled states of material particles
have been thoroughly studied in recent years both theoretically and
experimentally, and some schemes for their generation have been designed
and partially realized. Some studies concentrated on how
to produce entanglement between groups of two or few atoms,
exploiting for example the collective vibrational motion of trapped ions
\cite{molmer99,sackett2000}, the single-photon interference at photodetectors
\cite{cabrillo1999}, or the conditional dynamics of two atoms
within a single-mode cavity field \cite{plenio1999}.
More recently, there has been a growing interest on how to create
multipartite entanglement between atoms belonging to
a single atomic ensemble considered as a
multi-party quantum system,
by exploiting the interaction with a light field \cite{sorensen2001_1},
and the subsequent detection process \cite{duan2003,stockton2004}.
Finally, and more ambitiously, various schemes have been proposed for the entanglement
of different (two or more) macroscopic or mesoscopic atomic ensembles
\cite{duan2000,juls01,duan2002,ADL3,sherson05}.

In the cases of several (at least and typically two) macroscopic atomic ensembles,
where collective atomic operators can be described
by some continuous-variable approximation,
it is only possible to design schemes for the realization of weak entangled states.
Some of these schemes exploit quantum non-demolition (QND) measurements on
auxiliary electromagnetic fields (usually assumed in some Gaussian state)
interacting with the atoms to prepare
entangled states of atomic systems \cite{duan2000,juls01,ADL3}.
However, the probabilistic nature of the quantum measurement events makes
the generation of atomic entangled state conditioned by the measurement outcomes,
usually yielding a low probability of success.
This is particularly true for the preparation of maximally entangled states.
Such a shortcoming should be in principle overcome by exploiting the
knowledge of the state vector of the atomic system conditioned on the outcome of a measurement,
and then by introducing a proper feedback scheme to efficiently drive the system toward a
maximally entangled state. Actually, Stockton \emph{et al}. showed in Ref.~\cite{stockton2004}
how this strategy can be properly used to deterministically
prepare highly entangled Dicke states of \emph{a single} atomic ensemble.

In the present work we introduce a reliable feedback scheme to generate maximal entanglement
of \emph{two mesoscopic} atomic ensembles. In this scheme, the discrete quantum nature
of the atomic systems is fully taken into account \emph{without} resorting to any
continuous variable or Gaussian approximation.
Our proposal is based on the model introduced by Di Lisi and M{\o}lmer \cite{ADL3},
where two collections of atoms, probed by a sequence of single-photon scattering
processes, are conditionally entangled by QND measurements of the total atomic population
difference between the two atomic samples. This model has been recently shown to be
robust against spontaneous scattering \cite{DLDSI04}.

In the present work
the results of the QND measurements obtained by photo-detections are exploited
to drive the system into the maximally entangled state by a suitable feedback mechanism.
The feedback scheme that we introduce is a proper modification to
the fully discrete case of the continuous feedback strategy
originally designed by Thomsen, Mancini and Wiseman (TMW) \cite{TMW02})
to generate high spin squeezing of a single atomic ensemble, whose experimental realization was
recently obtained by Geremia \emph{et al}\cite{mabuchi}.
The same scheme was generalized to the case of two atomic ensembles
to produce two-mode spin squeezing \cite{BerrySanders02}.

Our procedure is monitored
by quantitative wave-function simulations which
show how the sequence of photo-detection events, followed by the feedback signal,
gradually modifies the state of the samples and post-selects the maximally entangled states.
We show that the feedback scheme enormously increases the rate of success
in producing maximally entangled states of the two atomic ensembles
compared with the scheme in which feedback is absent. We also show that
the efficiency is further improved by adiabatically
switching off the feedback signal; in this way one obtains
a quasi-deterministic generation of the maximally entangled state.
Finally, we study the problem for the more
realistic case of imperfect detectors, and we show how the feedback scheme
guarantees a very high probability of success in this case as well, making the
mechanism quite reliable against losses.

\section{The model}\label{sec:model}

\begin{figure}
\includegraphics[width=7cm,height=5cm]{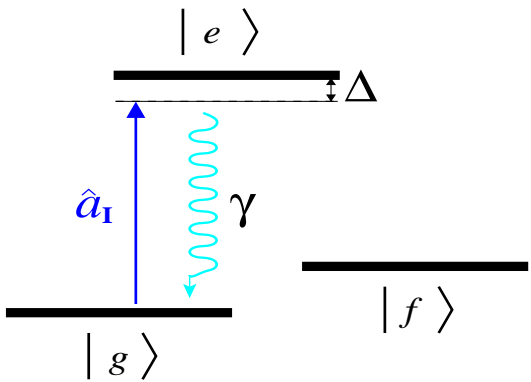}
\caption{\label{fig1} (Color online). Level structure of the atoms. $\ket{g}$ and $\ket{f}$ are metastable states,
$\ket{g}$ is coupled off resonantly by the electromagnetic field, whose annihilation
operator is $\an{a}_I$, to the excited state $\ket{e}$. Here $\gamma$ is the spontaneous transition rate
and $\Delta$ is the detuning between the coupling field and the atomic transition frequency.}
\end{figure}

The two atomic ensembles, denoted by ``$1$'' and ``$2$'' respectively,
are identical, and each one is constituted by $N$ identical atoms
in a static magnetic field,
whose level structure consists of two metastable lower states, $\ket{g}$ and $\ket{f}$,
that correspond to Zeeman sublevels of the
electronic ground state of alkali atoms,
and one excited state $\ket{e}$, cf. Fig~\ref{fig1}.

We then consider an optical beam passing through the atomic clouds which is coupled
(out of resonance, with detuning $\Delta$) only to the transition $\ket{g}\rightarrow\ket{e}$.
We can introduce the atomic spin operators for an atom $a$ in ensemble $i$ as
\bea
j^{(i)}_{a, x}&=&\frac{\ket{f}_a\!\bra{g} + \ket{g}_a\!\bra{f}}{2} \; , \nonumber \\
&& \nonumber \\
j^{(i)}_{a, y}&=&\frac{\ket{f}_a\!\bra{g} - \ket{g}_a\!\bra{f}}{2i} \; , \nonumber \\
&& \nonumber \\
j^{(i)}_{a, z}&=&\frac{\ket{f}_a\!\bra{f}-\ket{g}_a\!\bra{g}}{2} \; .
\eea
where $a=1,...,N$ is the atomic index and $i=1,2$ is the ensemble index.
The dynamics of the ensembles can be described by collective spin operators whose
$x$-,$y$- and $z$-components,
for each ensemble $i$, read
\be
\an{J}^{(i)}_{x}=\sum_{a=1}^{N}j^{(i)}_{a, x} \; , \; \;
\an{J}^{(i)}_{y}=\sum_{a=1}^{N}j^{(i)}_{a, y} \; , \; \;
\an{J}^{(i)}_{z}=\sum_{a=1}^{N}j^{(i)}_{a, z}\;.
\ee
These sums define the $x$-,$y$- and $z$-components of the collective angular momentum $\an{J}^{(i)}$.
In particular, the eigenvalues $m_i$ of $\an{J}^{(i)}_{z}$ are proportional to the population
difference in the two stable states.

\begin{figure}[tbp]
\includegraphics[width=8cm,height=6cm]{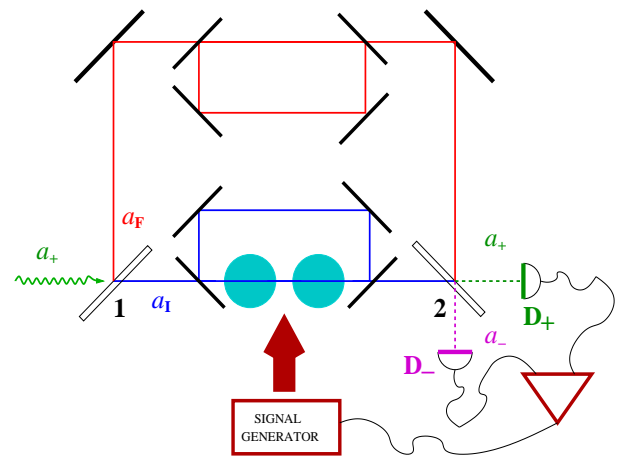}
\caption{\label{fig2} (Color online). Schematic experimental configuration.
Atoms occupying the internal state $\ket{g}$ in the two samples interact
with the light field, incident from the left.
The presence of the cavities guarantees a high
spectral and directional resolution of the entering photon.
The phase shift of the light field due to the interaction with the atoms
is registered by the different photocurrents in the two detectors. These signals
can be combined with that produced by a signal generator and hence sent back
to the two ensembles.}
\end{figure}
The two atomic ensembles are initially prepared, by optical pumping,
so that each ensemble is fully
polarized along the $x$-axis with collective spin equal to $J^{(i)}_{x}=N/2$
\cite{juls01,ADL3,TMW02,DLDSI04}. This means that the atoms are distributed between states $\ket{g}$ and $\ket{f}$
according to a binomial distribution with probability $1/2$ for each state.
The composed system made by
the two atomic clouds is described by the total spin operators
\be
\an{J}_{k}^{\pm}= \an{J}^{(1)}_{k}\pm \an{J}^{(2)}_{k}\;, \mbox{with}\; k=x,y,z \;.
\ee
The maximally entangled state of the composed system can be written as \cite{ADL3}
\be
\ket{\psi_{me}}= \frac{1}{\sqrt{N+1}}\sum_{m=-N/2}^{N/2}\ket{m}\otimes\ket{-m}\;.
\ee
Here $\ket{m_1}\otimes\ket{m_2}$ is the eigenstate of $\an{J}_{z}^{+}$
with eigenvalue $m_1+ m_2$. The state vector $\ket{\psi_{me}}$
is the only simultaneous eigenvector of $\an{J}_{z}^{+}$,
$\an{J}_{y}^{-}$ and $\an{J}_{x}^{-}$ with null eigenvalue
\cite{BerrySanders02,BerrySandersNJP02,hofmann2003}.
Obviously, the variances of these operators vanish
in the maximally entangled state $\ket{\psi_{me}}$ and, moreover, quite trivially,
\be\label{entcon}
\medial{\an{J}_{z}^{+}}_{me}=\medial{\an{J}_{y}^{-}}_{me}
=\medial{\an{J}_{x}^{-}}_{me}=0\;,
\ee
where, here and henceforth
$\medial{\an{A}}_j=Tr[\an{A}\varrho_j]$, where $\varrho_j$
is a generic density matrix, and
$\varrho_{me}=\ket{\psi_{me}}\bra{\psi_{me}}$ is the density
matrix of the maximally entangled state.
Eq. (\ref{entcon}) provides therefore \emph{a necessary} (but obviously not sufficient)
 condition for the realization
of the maximally entangled state and will be exploited in the following
to set up the feedback scheme.

The QND measurement of the total population difference between the two lower
atomic states by photo-detection
corresponds to the measurement of the observable $\an{J}_{z}^+$
(the explicit form of the QND interaction Hamiltonian is given by Eq.~(\ref{heff}) below),
and can be realized, for example, by using light with two polarization components
that interact differently with the two atomic states, generating in this way
different phase shifts that produce a polarization rotation signal. Another
method may consist in using modulated light, with one frequency component
closer to resonance than the other, so that the interaction with the atoms yields
a phase difference between the two components. However, for ease of presentation,
we follow the schematic model described in Ref.~\cite{ADL3}.

The two atomic ensembles are placed in one
arm of an interferometric setup, cf. Fig.~\ref{fig2}.
The incoming field, whose annihilation operator is denoted by $\an{a}_+$,
is a highly collimated single photon pulse, which
is decomposed in two components by means of a $50\%-50\%$ beam splitter:
the reflected component $\an{a}_{F}$ follows the free
path, while the transmitted component $\an{a}_{I}$ goes through
the atomic samples and interacts with them.

Since only the state $\ket{g}$ is
off-resonantly coupled with the excited level by the
interacting field, the phase shift between the two field components
is proportional to the population difference $m_1+m_2$ between
the atomic ground states \cite{ADL3}, and can be
resolved by the intensities measured in the two output ports of the interferometer
by photo-detector $D_+$, absorbing photons of mode
$\an{a}_+=(\an{a}_F+\an{a}_I)/\sqrt{2}$,
and photo-detector $D_-$, absorbing photons of mode $\an{a}_-=(\an{a}_F-\an{a}_I)/\sqrt{2}$.
The sequence of measurements of the field phase shift
corresponds to QND measurements of the
observable $\an{J}_{z}^{+}$ and yields a nondestructive evolution of
the global state of the two atomic ensembles \cite{ADL3}.

In order to guarantee a high spectral and directional resolution
of the entering photon, we can place in each arm of
the interferometer two symmetric ring cavities, in one of which the two atomic samples are
enclosed (see Fig.~2).
The ring cavity in the ``empty'' arm of the interferometer is needed to achieve optimal mode matching
at the output beam splitter of the interferometer.
This, together with the condition of a
detuning $\Delta$ much larger than the excited level decay rate $\gamma$,
greatly reduces the effect of spontaneous emission which can
then be omitted from our analysis. Moreover, by considering a traveling-wave probe
in this far-off resonance case, any information about the
relative positions of the atoms of the ensembles due to recoil is erased.
Hence, the effective Hamiltonian of the atom-photon
interaction can be written as \cite{ADL3}
\be\label{heff}
H_{eff} = \frac{g^2}{\Delta}(N -\an{J}_{z}^{+})\ad{a}_{I}\an{a}_{I}\;,
\ee
where $g$ is the coupling strength between
the single atom and the radiation field.

The modification of the state of the two samples induced by
the photon-atom interaction and the subsequent QND measurement
can be formalized by the action of an appropriate POVM \cite{peres}.
Following Ref.~\cite{ADL3}, we assume that
photodetectors $D_+$ and $D_-$ are able to
reveal one photon at a time.
Therefore, the state of the two atomic ensembles after each photo-detection
is determined by the action of the (Kraus) operators
$\an{M}_{\pm}=(\an{I}\pm e^{-i\chi(N-\an{J}_{z}^{+})})/2$,
where $\an{I}$ is the identity operator and $\chi=g^2\tau/\Delta$
is the phase shift, ($\tau$ is the duration of the photon-atomic ensembles
interaction process).

The dynamics induced by the POVM,
which preserves the rotational symmetry of the atomic system, and hence the value of the
total angular momentum, cannot produce an evolution towards
a maximally entangled state, which is a linear combination of eigenstates
of different total spin $\an{J}=\an{J}_1+\an{J}_2$.
The rotational symmetry can be broken by rotating,
with opposite angles $\Omega$ and $-\Omega$, each atomic sample around the $x-$axes
in the spin space.
Actually, such a rotation is
currently used in experiments for purely practical reasons,
for example to clean up the relevant signal
from technical noise at high frequency \cite{juls01}.
The rotation is performed after each photo-detection,
and is realized by the operator $U_R=\exp[-i\Omega J_{x1}]\exp[i\Omega J_{x2}]=\exp[-i\Omega\an{J}_{x}^{-}]$.
In this way what is effectively
measured at the $n$-th photo-detection is the rotated operator
\bea\label{jzrot}
\an{J}_{z}^{+}(n\Omega ) &=&
e^{i n\Omega \an{J}_{x}^{-}} \an{J}_{z}^{+} e^{-i n\Omega\an{J}_{x}^{-}}\nonumber\\
&=& (J_{z1}+J_{z2})\cos(n\Omega )\nonumber\\
&&+(J_{y1}-J_{y2})\sin(n\Omega)\;,
\eea
as it appears in the non-rotated frame.

Photo-detection losses are accounted for by considering
a finite efficiency of the measurement process, i.e. assuming that only a fraction $\eta < 1$
of the probe photons is actually detected. In this non-ideal situation
the evolution of the density matrix is timed by the rate
at which the single photon enters the interferometric set-up, and is
conditioned by the possibility of photo-detection.

If $\varrho_{n}$ is the density matrix of the total atomic
spin system after $n$ photons were sent on it,
the (non-normalized) state
$\tilde{\varrho}^{\prime}_{n+1}$ at the successive step
is given by
\be\label{up_st_pmd}
\tilde{\varrho}^{\prime}_{n+1}=
e^{-i\Omega\an{J}_{x}^{-}}\an{M}_{\pm}\varrho_{n}\ad{M}_{\pm}
e^{i\Omega\an{J}_{x}^{-}}
\ee
if the photon is detected by detector $D_+$ ($\an{M}_{+}$), or $D_-$ ($\an{M}_{-}$),
with probability
$P_{\pm}=\eta Tr[\an{M}_{\pm}\varrho_{n}\ad{M}_{\pm}]$ respectively;
otherwise, if, with probability $1-\eta$, no photo-detection occurs,
it can be written as
\be\label{up_st_nd}
\tilde{\varrho}^{\prime}_{n+1}=
e^{-i\Omega\an{J}_{x}^{-}}\left(\an{M}_{+}\varrho_{n}\ad{M}_{+} +
\an{M}_-\varrho_{n}\ad{M}_-\right)e^{i\Omega\an{J}_{x}^{-}}\;.
\ee

\section{The feedback scheme}

Our purpose is to efficiently realize the maximally entangled state $\varrho_{me}$
in a controlled fashion. As shown in Ref.~\cite{ADL3},
the measurement scheme described in section \ref{sec:model} yields only a
certain probability that the global state of
the two atomic ensembles will be gradually projected onto $\varrho_{me}$.
Moreover, we recall that, to this aim, Eq.~(\ref{entcon}) is only
a necessary condition and, although in the initial state $\varrho_0$ we have
$\medial{\an{J}_{z}^{+}}_0=\medial{\an{J}_{y}^{-}}_0=\medial{\an{J}_{x}^{-}}_0=0$,
as the measurements process goes on, the state of the two samples does not necessarily
satisfy this condition any longer. In fact, since we are measuring $\an{J}_{z}^{+}$,
the back-action of the POVM has the effect of
decreasing the uncertainty of $\an{J}_{z}^{+}$ and, at the same time,
randomly shifting $\medial{\an{J}_{z}^{+}}$ from its initial null value~\cite{ADL3,TMW02}.
As shown in Ref.~\cite{TMW02},
this shift can be interpreted as a stochastic rotation of the mean
collective atomic spin around the $y$-axis induced by the measurement process.
The optimal situation we would like to require is
$\medial{\an{J}_{z}^{+}}_n=\medial{\an{J}_{y}^{-}}_n=\medial{\an{J}_{x}^{-}}_n=0$
at each step of the measurement process. We choose this requirement because
it forces the state of the
two atomic ensembles to satisfy condition (\ref{entcon}) at each stage of its
conditional evolution and, due to the fact that
$\ket{\psi_{me}}$ is the only simultaneous eigenvector of $\an{J}_{z}^{+}$, $\an{J}_{y}^{-}$
and $\an{J}_{x}^{-}$ with null eigenvalue, this procedure
likely increases the probability to drive the system into the maximally entangled state.
In fact, the validity of this choice is verified {\it a posteriori} by observing
that the variances of the three collective operators $\an{J}_{z}^{+}$, $\an{J}_{y}^{-}$
and $\an{J}_{x}^{-}$ are monotonically decreasing functions during the time
evolution.
Therefore, to implement the optimal condition, Eq.~(\ref{entcon}), we impose a quantum
feedback that properly counter-rotates the atomic spin operators.
From the expression (\ref{jzrot}) of the rotated operator
at $n$-th step and photo-detection,
it follows that the optimal condition on
$\an{J}_{z}^{+}$ and $\an{J}_{y}^{-}$ is obtained by imposing
$\medial{\an{J}_{z}^{+}(n\Omega)}_n=0$.
The simplest way to achieve this situation is to act with a
unitary feedback operator $\an{U}_f=\exp{[i\lambda\an{J}_{y}^{+}]}$ \cite{TMW02}.
This operator realizes a rotation of an angle $\lambda$ around the
instantaneous
$y-$axes in the non-rotated frame and acts simultaneously on both
atomic ensembles after the photon has been detected.
This rotation can be realized, for example, by applying
the combination of a static and an amplitude controlled rf magnetic field, which
couples the two ground states $\ket{g}$ and $\ket{f}$ and drives the $y$-components
of the collective spin operators generating the rotation
$\an{U}_f=\exp{[i\lambda\an{J}_{y}^{+}]}$ \cite{TMW02}.
However, in our scheme, the amplitude of the driving field is controlled by the
feedback parameter $\lambda$, that
depends upon the measured signal.
The state of the system after the photo-detection,
the rotation around the ``fixed'' $x$-axes, and the action of the feedback, is
given by the controlling updating equation
\be
\tilde{\varrho}_{n+1}=
e^{i\lambda\an{J}_{y}^{+}}e^{-i\Omega\an{J}_{x}^{-}}\an{M}_{\pm}\varrho_{n}\ad{M}_{\pm}
e^{i\Omega\an{J}_{x}^{-}}e^{-i\lambda\an{J}_{y}^{+}}\;.
\ee
Obviously, if no detection occurs any feedback signal is implemented,
and the updated state is given by Eq.~(\ref{up_st_nd}).

Imposing the condition
$Tr\left[\an{J}_{z}^{+}\tilde{\varrho}_{n+1}\right]=0$, we get
the required form of the feedback parameter $\lambda$ ($\lambda_{\pm}$), expressed by
the relation:
\begin{widetext}
\be\label{fc}
\tan\lambda_{\pm} =-\frac{\medial{\an{J}_{z}^{+}(\Omega)}_n+
\medial{e^{-i\chi\an{J}_{z}^{+}}\an{J}_{z}^{+}(\Omega)e^{i\chi\an{J}_{z}^{+}}}_n
\pm\medial{\an{J}_{z}^{+}(\Omega)e^{-i\chi(N-\an{J}_{z}^{+})}+
e^{i\chi(N-\an{J}_{z}^{+})}\an{J}_{z}^{+}(\Omega)}_n}
{\medial{\an{J}_{x}^{+}}_n+
\medial{e^{-i\chi\an{J}_{z}^{+}(n)}\an{J}_{x}^{+}e^{i\chi\an{J}_{z}^{+}(n)}}_n
\pm\medial{\an{J}_{x}^{+}e^{-i\chi[N-\an{J}_{z}^{+}(n)]}+
e^{i\chi[N-\an{J}_{z}^{+}(n)]}\an{J}_{x}^{+}}_n}\;.
\ee
\end{widetext}

The phase shift $\chi$ is typically very small and
when $\chi N \ll 1$ (which is easily satisfied for $N\simeq 100$),
we can expand the numerator and the
denominator of $\tan\lambda_{\pm}$ to second order in $\chi$.
Since in the non-rotated frame the action of the feedback imposes
$\medial{\an{J}_{z}^{+}(n\Omega)}_n=0$ at \emph{each} step of the photo-detection,
Eq.~(\ref{jzrot}) implies that $\medial{\an{J}_{z}^{+}}_n=\medial{\an{J}_{y}^{-}}_n=0$, and,
as long as our scheme holds, $\medial{\an{J}_{x}^{-}}_n=0$ is naturally satisfied.
Retaining the leading terms in $\chi N$,
we obtain the explicit expressions of the
feedback parameter for the two
possible photo-detection outcomes:
\bea
\tan\lambda_+ & = & \frac{ -\chi^2N(2\medial{(\an{J}_{z}^{+})^2}_n\cos\Omega+
\medial{[\an{J}_{y}^{-},\an{J}_{z}^{+}]_+}_n\sin\Omega)}
{ 4\medial{\an{J}_{x}^{+}}_n}, \nonumber \\
\tan\lambda_- & = & \frac{ 2\medial{(\an{J}_{z}^{+})^2}_n\cos\Omega+
\medial{[\an{J}_{y}^{-},\an{J}_{z}^{+}]_+}_n\sin\Omega}
{ N\medial{\an{J}_{x}^{+}}_n}\; . \label{lambdapm}
\eea
with $[\an{A},\an{B}]_+=\an{A}\an{B}+\an{B}\an{A}$.

\section{Numerical simulation and the adiabatic feedback}

We can now test by numerical simulations the measurement and feedback schemes.
We simulate the conditional evolution of the global state of the two atomic
samples, due to the acquisition of information by successive single photodetections,
according to the updating equations for $\tilde{\varrho}_{n+1}$,
where the angle of the feedback rotation
is determined step by step either by
the first or by the second of Eqs.~(\ref{lambdapm}).

For each case we have simulated 50 evolutions, with $5\times 10^4\div 10^5$
incoming photons. For every simulation we compute the overlap with the maximally entangled state
$Tr[\varrho_n\varrho_{me}]$ and
evaluate the degree of entanglement between the two atomic ensembles.
Since in the case of perfect detection ($\eta=1$) we always deal with
pure states, the entanglement is quantified by
the von Neumann entropy $E=-Tr[\varrho_j\log_2 \varrho_j]$
associated with the reduced density matrices of the
two subsystems of $N$
atoms, $\varrho_j$. If we restrict ourselves to states which are
symmetric under permutations inside each subensemble,
$E$ takes values between zero for a product state, and
$\log_2(N+1)$ for a maximally entangled state of the two samples.

If $\eta\neq 1$ (mixed states) the entanglement can be efficiently quantified
by \emph{the relative violation of the local uncertainty relations (LUR) for the relevant observables
$\an{J}_{z}^{+}$, $\an{J}_{y}^{-}$ and
$\an{J}_{x}^{-}$}, as shown by Hofmann and Takeuchi~\cite{hofmann2003}
for entanglement generation in ($N+1$)-level systems.
This quantity is defined
as $C_{LUR}=1-(\delta\an{J}_{z}^{+}+\delta\an{J}_{y}^{-}+\delta\an{J}_{x}^{-})/N$,
where $\delta\an{A}$ denotes the variance of $\an{A}$ and $C_{LUR}=1$
for maximally entangled states.

We quantify the efficiency of the scheme by determining the fraction of simulations $\mathcal{F}$ for which
the value of the overlap between the final state of
the two atomic ensembles after the photo-detection sequence and
the maximally entangled state is larger than $0.99$.
The results of the simulations show a substantial efficiency of the feedback
scheme (FS) compared with those obtained by pure probabilistic schemes (PS)
without feedback.
\begin{figure}[tbp]
\includegraphics[width=8cm,height=6.5cm]{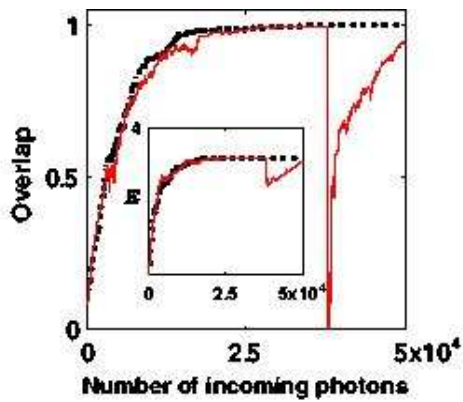}
\caption{\label{fig3}(Color online).
Numerical simulations of the evolution in the presence of feedback of
the overlap ($Tr[\varrho_n\varrho_{me}]$) between the state of the two atomic samples ($N=10$)
and the maximal entangled state in a case in which the feedback scheme is successful ({\it S}, dotted line) in generating
$\varrho_{me}$ and in a case in which the scheme is broken ({\it B}, solid line) by ``anomalous''
feedback angles (see text); inset: Evolution of the entanglement $E$ in the {\it S} case
(dotted line) and in the {\it B} case (solid line).
The values of the parameters used in the simulations are:
$\eta=1$, $\chi=0.03$ (achievable with highly collimated beams),
$\Omega=\pi/10$ and the total number of incoming photons is $n_{ph}=5\times10^4$.}
\end{figure}
\begin{figure}[tbp]
\includegraphics[width=8.7cm,height=5.5cm]{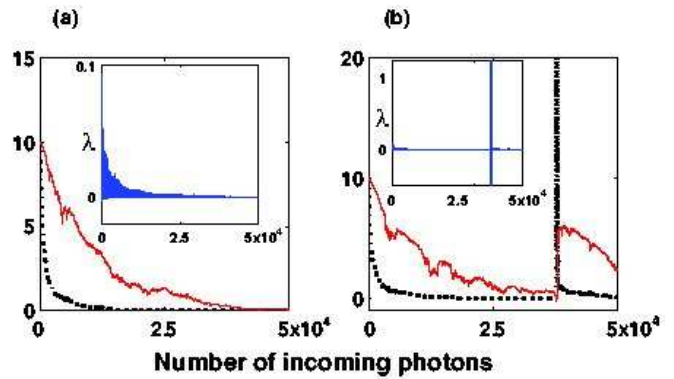}
\caption{\label{fig4}(Color online).
(a): Numerical simulations of the evolution of
$\medial{\an{J}_{x}^{+}}$ (solid line) and of $\delta\an{J}_{z}^{+}+\delta\an{J}_{y}^{-}$ (dotted line)
 in the {\it S} case; inset (a): evolution of the feedback angle $\lambda$ during the {\it S} simulation;
(b): Evolution of $\medial{\an{J}_{x}^{+}}$ (solid line) and of
$\delta\an{J}_{z}^{+}+\delta\an{J}_{y}^{-}$ (dotted line)
in the {\it B} case. Notice that after the point in which
$\medial{\an{J}_{x}^{+}}<\delta\an{J}_{z}^{+}+\delta\an{J}_{y}^{-}$,
the variances \emph{explodes}. This corresponds
to \emph{exploding} values of $\lambda$ around that point (inset (b)).
Parameter values are the same as in Fig.~3.}
\end{figure}
\begin{figure}[tbp]
\includegraphics[width=6cm,height=5.5cm]{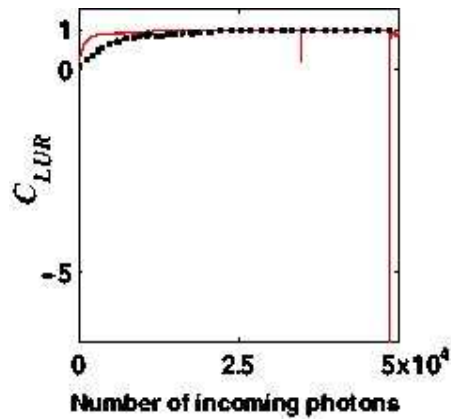}
\caption{\label{fig5} (Color online).
Evolution of $C_{LUR}$ in a {\it S} simulation (dotted line)
and in a {\it B} simulation (solid line). Positive values of $C_{LUR}$
provide a quantitative estimate of the amount of entanglement.
Parameter values are the same as in Fig.~3 except that now $\eta=0.9$.}
\end{figure}
\begin{figure}[tbp]
\includegraphics[width=8.7cm,height=5cm]{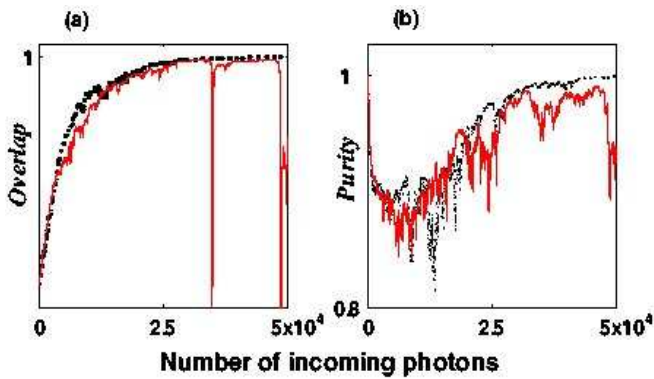}
\caption{\label{fig6} (Color online).
(a): Evolution of the overlap
in the {\it S} case (dotted line) and in the {\it B} case (solid line);
(b): Evolution of the purity $Tr[\varrho_{n}^2]$ in the {\it S} case (dotted line)
and in the {\it B} case (solid line). Notice that a successful feedback scheme
has also a purification action. Obviously when the scheme is broken the purification effect is lost.
Parameter values are the same as in Fig.~5.}
\end{figure}
We now discuss the different cases:
\begin{itemize}
\item {\it a) No feedback}: in this (PS) case the successful rate is very low: in fact, we have
$\mathcal{F} \cong 6\%$ both in the ideal case ($\eta = 1$) and in the
more realistic case ($\eta = 0.9$).
\item {\it b) Simple feedback}: the action of the simple feedback
relevantly increases the successful rate with respect to the pure PS. In fact,
in the ideal case of perfect detection we have $\mathcal{F} \approx 54\%$:
the action of the feedback on the mean value and the variances of the
relevant spin operators $\an{J}_{z}^{+}$, $\an{J}_{y}^{-}$ and
$\an{J}_{x}^{-}$ is extremely effective, driving them to the ideal null values
as the state of the two atomic clouds ($\varrho_n$) approaches to the maximally
entangled state
($\varrho_{me}$), cf. Fig.~\ref{fig4}(a) and \ref{fig4}(b).
In the more realistic case ($\eta = 0.9$) we obtain $\mathcal{F} \approx 24\%$.
The worse result in this latter case is obviously
due to the the fact that the instability of the scheme
is even stressed by the mixedness of the state of the
two atomic ensembles, making the action of the feedback
less ``precise'' than the case with pure state.
Moreover, the evolution of the entanglement is
slower, requiring many more photons to be sent on
the samples to get the same values of the overlap
of the case with $\eta=1$, cf. Figs.~\ref{fig5}~and~\ref{fig6}.
It is remarkable that,
since the feedback signal depends only on
the statistical moments of the spin observables,
its efficiency does not depend on the exact values
taken by the feedback angles. In fact,
extended numerical simulations show that
the feedback mechanism is not appreciably
affected by perturbing the values of $\lambda$ by
fluctuations of the order of
$\lambda_{\pm}\times 10^{-1}\div 10^{-2}$.
This is an important aspect of our scheme, which is therefore
quite robust against imperfections in the actuation of the feedback operation.
\item {\it c) Modified (``adiabatic'') feedback}:
The only partial success of the feedback scheme is caused by
the possibility that Eqs.~(\ref{lambdapm})
provide wrong values of the feedback angle when $\varrho_n\approx\varrho_{me}$.
In fact, in this situation, the feedback angle $\lambda$ should uniformly go to zero,
otherwise, since $\varrho_{me}$ is not invariant under the feedback rotation, the state of the two
samples is transformed in a state practically orthogonal to the maximally entangled one,
cf. Fig.~\ref{fig3}.
This case could happen when $\medial{\an{J}_{x}^{+}}$
becomes smaller than the numerators in
Eqs.~(\ref{lambdapm}),
essentially proportional to
$\delta\an{J}_{z}^{+}+\delta\an{J}_{y}^{-}$, cf. Fig.~\ref{fig4}~(b).
Actually, from the uncertainty relations for spin systems is easy to show that
$\delta\an{J}_{z}^{+}\delta\an{J}_{y}^{-}\rightarrow 0 $
$\Rightarrow$ $\medial{\an{J}_{x}^{+}}\rightarrow 0$ \cite{BerrySanders02,BerrySandersNJP02},
and, due to the random jump caused by
the back action of the photo-detection, in this situation it is possible to have
$\medial{\an{J}_{x}^{+}}<\delta\an{J}_{z}^{+}+\delta\an{J}_{y}^{-}$, getting in this way
``exploding'' values for $\lambda$, cf. inset Fig.~\ref{fig4}~(b).
However, it is possible to avoid these ``anomalous'' feedback angles
by modifying the feedback algorithm in order to force $\lambda_{\pm}$ to go uniformly to zero
as the maximally entangled state is approached.
In fact, inspired by the mean values taken by $\lambda{\pm}$
in simulations in which the maximal entangled state is reached (inset~Fig~\ref{fig4}(a)),
we can impose on the feedback angles of Eqs.~(\ref{lambdapm})
a decreasing exponential cut, a sort of adiabatic switching off,
expressed as a function of the number $n$ of incoming
photons: more precisely, the exponential cut is $\lambda_{cut}=x\exp(-x)$,
with $ x \equiv n \times 10^{-4}$, and
we impose that if $|\lambda_{\pm}|\geq \lambda_{cut}$, then
$\lambda=\varepsilon(\lambda_{\pm})\lambda_{cut}$,
where $\varepsilon$ is the signum function; otherwise $\lambda=\lambda_{\pm}$.
However, in order to not reduce the effectiveness of the feedback scheme, we switch on
the correction procedure only after a high degree of entanglement is obtained,
so that the state of the atomic ensembles is ``close'' to the maximally entangled one.
The point in which the correction procedure should begin to act
has to be chosen in such a way that the probability that anomalous values of
$\lambda$ have already occurred is very low:
by a statistical analysis of the simple feedback scheme, this point coincides roughly with
$n\approx 2\times 10^4$. In this way we can softly, even though more slowly,
drive the state of the system
toward the maximally entangled state.
With this procedure, the results of the numerical simulations are remarkable. In the case
with $\eta=1$, we have $\mathcal{F} \approx 92\%$, and the successful rate
remain remarkably high also for the realistic choice $\eta=0.9$: $\mathcal{F} \approx 78\%$.
The reliability of the scheme can also be appreciated by considering stronger constrained data:
in fact, if we consider overlaps greater than $0.999$, we have $90 \%$ for $\eta=1$ and
$64 \%$ for $\eta=0.9$, while overlaps greater than $0.9999$ (practically, a full
realization of the maximally entangled state) leads to $80 \%$ for $\eta=1$ and $50 \%$
for $\eta=0.9$.
\end{itemize}

\section{Conclusions}

In conclusion, we have introduced an experimentally
feasible, quasi-deterministic stroboscopic feedback scheme able
to generate the maximally
entangled state of a system composed by two mesoscopic atomic ensembles,
each containing a number of atoms $N \sim 10-10^3$.

The feedback scheme is based on QND measurements of the
total atomic population difference between the internal states
of the two atomic samples, which are probed by a photon beam.
Though the state of the atomic system is conditionally
entangled by the photodetections, the conditional evolution
does not necessarily lead to the the maximally entangled state.
To this aim, we designed a scheme in which the signal obtained
by the measurements, and the corresponding information on the state of the
ensembles, is exploited to set up a feedback signal, formalized by
an unitary operator acting on the state of the atomic samples, that
drives the system into the maximal entangled state. The feedback strategy adopted
is a modification of that introduced in Ref.~\cite{TMW02} by Thomsen, Mancini and Wiseman.

The procedure is analyzed by numerical simulations of the conditional
evolution of the state of the two atomic samples. In particular, for every simulation, we have
monitored the \emph{overlap} between the state of the atomic system and
the maximally entangled state, and we evaluated the degree
of entanglement between the two atomic ensembles. The efficiency of the protocol
is quantified by the fraction of simulations in which the final value of the \emph{overlap}
is larger than $0.99$.

The results of the simulations show that the implementation
of the feedback scheme yields a considerable enhancement of the efficiency
of generating the maximal entangled state of the two atomic ensembles,
compared with the case of pure probabilistic schemes without feedback.
Moreover, we have shown that this efficiency
can be further improved by modifying the procedure by means of an adiabatic
switching off of the feedback signal.

We have shown that the scheme is able to tolerate a number of experimental imperfections,
as, for instance, a limited detection efficiency. In the numerical simulations of the feedback
protocol, we have considered a photocounting efficiency $\eta=0.9$: although this value is optimistic
(typical laboratory photocounters are characterized by $\eta\sim 0.5$),
the impressive, continuing progresses in this field give hope that very high efficiencies will be
available in the near future
(see for instance Ref.~\cite{yama}, where $\eta \simeq 0.88$ has been achieved at a particular wavelength). Nevertheless,
at variance with the experimentally achieved \cite{juls01}
probabilistic generation of weakly entangled states between macroscopic
atomic ensembles (with typical numbers of atoms $N \sim 10^{13}$),
our quasi-deterministic scheme applied to
mesoscopic ensembles allows to access a presently
unexplored area of entanglement production in atomic systems, even with present-day
technologies against photon losses and imperfect detections.

\end{document}